\documentclass[12pt]{article}
\usepackage{latexsym,amsmath}
\usepackage{graphicx}

\def\beq{\begin{eqnarray}}    
\def\eeq{\end{eqnarray}}      
\def\at{\left(}               
\def\aq{\left[}               
\def\ct{\right)}              
\def\cq{\right]}              

\newcommand{\be}{\begin{equation}}
\newcommand{\ee}{\end{equation}}
\newcommand{\bea}{\begin{eqnarray}}
\newcommand{\eea}{\end{eqnarray}}
\newcommand{\beaa}{\begin{eqnarray*}}
\newcommand{\eeaa}{\end{eqnarray*}}

\newcommand{\nn}{\nonumber \\}

\begin{document}
\begin{center}
VANISHING COSMOLOGICAL CONSTANT IN MODIFIED GAUSS - BONNET GRAVITY WITH CONFORMAL ANOMALY   \\
\bigskip
Iver Brevik\footnote{E-mail:  iver.h.brevik@ntnu.no}\\
Department of Energy and Process Engineering\\
Norwegian University of Science and Technology\\
N-7491 Trondheim, Norway\\

\bigskip

John Quiroga Hurtado\footnote{E-mail: jquiroga@utp.edu.co}\\
Department of Physics\\
Universidad Tecnol\'ogica de Pereira\\
Colombia\\

\bigskip
\begin{center}
October 2006\\
\end{center}

\vfill

{\bf Abstract}

\end{center}

We consider dark energy cosmology in a de Sitter universe filled
with quantum conformal matter. Our model represents a Gauss-Bonnet
model of gravity with contributions from quantum effects. To the
General Relativity action an arbitrary function of the GB
invariant, $f(G)$, is added, and taking into account quantum
effects from matter the cosmological constant is studied. For the
considered model the conditions for a vanishing cosmological
constant are considered. Creation of a de Sitter universe by
quantum effects in a GB modified gravity is discussed.

\vfill


\newpage
There is a growing interest towards various studies of the
reported acceleration of  our observable Universe \cite{SuNv}.
This is motivated by recent astrophysical data analyses, which
hint to such a behavior. In order to explain it, the simplest
possibility is to introduce a dark energy component, whose origin
remains however uncertain.

The existence of such an energy, with almost uniform density
distribution and a substantial negative pressure which completely
dominates all other forms of matter, is inferred from recent
astronomical observations \cite{SuNv}. In particular, according to
recent astrophysical analysis, this dark energy seems to behave
like a cosmological constant, and  is responsible for the
accelerating expansion of the  universe. And there are reasons to
believe that the answer to this question has much to do with the
possibility to explain the physics of the very early Universe.

Models of dark energy are abundant. One of the proposed candidates
for it is the phantom,  called so because it implies a negative
energy field. The peculiar properties of a phantom scalar (with
negative kinetic energy) in a space with non-zero cosmological
constant have been discussed in an interesting paper by Gibbons
\cite{Gibbons}. As indicated there,  phantom properties bear some
similarity with quantum effects \cite{phtmtr}.
 An important property of the
investigation in \cite{Gibbons} is that it is easily generalizable
to other constant curvature spaces, such as the Anti-de Sitter
(AdS) space. There is presently considerable interest in such
spaces, coming in particular from the AdS/CFT correspondence.
According to it, the AdS space may have  cosmological relevance
\cite{cvetic}, e.g. by  increasing the number of particles created
on a given subspace \cite{jcap}. It could also be used to study a
cosmological AdS/CFT correspondence \cite{b494}: the study of a
phantom field in AdS space may give us a hint about the origin of
such a field via the dual description. In the supergravity
formulation, one may think of the phantom as  a special RG flow
for scalars in gauged AdS supergravity. (Actually, such an RG flow
may correspond to an imaginary scalar.)

Another candidate for dark energy is the tachyon \cite{EEQJ1,
snsdo}. This is an unstable field. The interest of models
exhibiting a tachyon is motivated by its role in the
Dirac-Born-Infeld (DBI) action as a description of the D-brane
action \cite{copeland,s1,s2}. In spite of the fact that the
tachyon represents an unstable field, its role in cosmology is
still considered useful as a source of dark matter
\cite{snsdo,Gibbons2} and, depending on the form of the associated
potential \cite{Paddy,Bagla,AF,AL,GZ}, it can lead to a period of
inflation. On the other hand, it is important to realize that a
tachyon with negative kinetic energy (yet another type of phantom)
can be introduced \cite{hao}. In that phantom/tachyon model the
thermodynamical parameter $w$ is naturally negative. In this case
the late time de Sitter attractor solution is admissible, and this
is one of the main reasons why it can be considered as an
interesting model for the dark energy \cite{hao}. Moreover, in
order to understand the role of the tachyon in cosmology it is
necessary to study its effects on other backgrounds, as in the
case of an anti-de Sitter background \cite{EEQJ1}.

On the other hand the origin of the so-called dark energy could be
related with the problem of the cosmological constant. One of the
most interesting approaches to this question is the modified
gravity. Actually, it is not absolutely clear why standard General
Relativity should be trusted at large cosmological scales. One may
assume that, considering minimal modifications, the gravitational
action  contains some additional terms growing slowly in the
presence of decreasing curvature \cite{capozziello, NOPRD, ln, tr,
string} (for a review, see \cite{GB2}), which could be responsible
for the current acceleration. In such a scenario one of the most
accepted approaches is the model of Modified Gauss-Bonnet Gravity.
In these models these additional terms in the gravitational action
are introduced by adding to the action a function depending on the
scalar curvature $R$ and the Gauss-Bonnet invariant $G$
\cite{GB1}. In this way it is possible to demonstrate that such
models lead to a plausible effective cosmological constant,
quintessence, or a phantom era. From these results  one may
conclude that as concerns the role of a gravitational alternative
for DE, the  modified GB gravity may be a very important candidate
\cite{GB-gravity}.

In the present paper we consider a Gauss-Bonnet model of gravity
with contributions from quantum effects. To the General Relativity
action an arbitrary function of the GB invariant, $f(G)$, is
added. Taking into account quantum effects from matter the
cosmological constant is studied. The conditions for vanishing of
the cosmological constant are studied, and their effects on the
stability of the de Sitter universe via quantum effects are
discussed.

 Let us begin with the action for the modified
gravity in the following form \cite{GB-gravity}:

\be \label{GR1} S=\int d^4 x\sqrt{-g}\left(F(G,R) + {\cal
L}_m\right)\ . \ee

Here $G$ is the GB invariant:
\begin{equation}
G=R^2 -4 R_{\mu\nu} R^{\mu\nu} +
R_{\mu\nu\xi\sigma}R^{\mu\nu\xi\sigma}\,. \label{GB}
\end{equation}
Varying over $g_{\mu\nu}$ we get

\bea \label{GR2} && 0= T^{\mu\nu} + \frac{1}{2}g^{\mu\nu} F(G,R)
\nn && -2 F_G(G,R) R R^{\mu\nu} + 4F_G(G,R)R^\mu_{\ \rho}
R^{\nu\rho} \nn && -2 F_G(G,R) R^{\mu\rho\sigma\tau}R^\nu_{\
\rho\sigma\tau}
    -4 F_G(G,R) R^{\mu\rho\sigma\nu}R_{\rho\sigma} \nn
&& + 2 \left( \nabla^\mu \nabla^\nu F_G(G,R)\right)R
    - 2 g^{\mu\nu} \left( \nabla^2 F_G(G,R)\right)R \nn
&& - 4 \left( \nabla_\rho \nabla^\mu F_G(G,R)\right)R^{\nu\rho}
    - 4 \left( \nabla_\rho \nabla^\nu F_G(G,R)\right)R^{\mu\rho} \nn
&& + 4 \left( \nabla^2 F_G(G,R) \right)R^{\mu\nu} + 4g^{\mu\nu}
\left( \nabla_{\rho} \nabla_\sigma F_G(G,R) \right) R^{\rho\sigma}
\nn && - 4 \left(\nabla_\rho \nabla_\sigma F_G(G,R) \right)
R^{\mu\rho\nu\sigma} \nn && - F_R(G,R) R^{\mu\nu} + \nabla^\mu
\nabla^\nu F_R(G,R) \nn && - g^{\mu\nu}\nabla^2 F_R(G,R) \ , \eea
$T^{\mu\nu}$ being the matter-energy momentum tensor, and where
the following expressions are used: \be \label{GR3}
F_G(G,R)=\frac{\partial F(G,R)}{\partial G}\ ,\quad
F_R(G,R)=\frac{\partial F(G,R)}{\partial R}\ . \ee The
spatially-flat FRW universe metric is chosen as \be \label{FRW}
ds^2=-dt^2 + a(t)^2 \sum_{i=1}^3  \left(dx^i\right)^2\ . \ee

Quantum effects will be taken into account by including the
contributions from the conformal anomaly
\begin{eqnarray}
T=b\left( F+{2\over 3}\,\nabla^2 R\right)+b'G+b''\,\nabla^2 R \,,
\label{anom}
\end{eqnarray}

where $T$ is the trace of $T^{\mu\nu}$ and $F$ is the square of
4-D Weyl tensor,

\be
F=R_{\mu\nu\alpha\beta}R^{\mu\nu\alpha\beta}-2R_{\mu\nu}R^{\mu\nu}
+{1\over 3}\,R^2\;. \ee

Note that such a conformal anomaly may be related to a bulk de
Sitter space (see \cite{SDOSN}).

Taking the trace of (\ref{GR2}), we get the following equation

\be\label{GR10}
0=T+2F(G,R)-\frac{1}{3}R^2F_G(G,R)-R(\nabla^2F_G(G,R))-3\nabla^2F_R(G,R))-RF_R(G,R).\ee

We are interested in the de Sitter type solutions where the Ricci
scalar, the Gauss-Bonnet invariant, and the square of 4-D Weyl
tensor  are constants:

\be R = R_0 \qquad G=G_0=\frac{1}{6}R_0^2\qquad F=0 \,.\ee

Assuming the maximally symmetric metric solution, we get

\be\label{GR11}
0=T+2F(G_0,R_0)-\frac{1}{3}R_0^2F_{G_0}(G_0,R_0)-R(\nabla^2F_{G_0}(G_0,R_0))-3\nabla^2F_R(G_0,R_0))-R_0F_R(G_0,R_0).\ee
For the CA we get

\be\label{GR12} T=b'G_0\,,\ee

and  Eq.(\ref{GR11}) takes the form

\be\label{GR13}
\frac{1}{2}R_0F_R(G_0,R_0)=F(G_0,R_0)-G_0F_{G_0}(G_0,R_0)+b'G_0.\ee

As an example let us consider case when $F(G,R)=R+f(G)$, (here $2
\kappa^2=16\pi G=1$); then one has

\beq G_0 f'(G_0)- f(G_0)-b'G_0=\frac{R_0}{2}\,. \label{exG1} \eeq

In general, solving Eq. (\ref{GR13}) in terms of $R_0$, one can
rewrite the maximally symmetric  solution as \beq
R^0_{\mu\nu}=\frac{R_0}{4}g^0_{\mu\nu}=\Lambda_{eff}\,g^0_{\mu\nu}\,,
\eeq which defines an effective cosmological constant. For the
considered example, when $F(G,R)=R+f(G)$, one has

\beq \Lambda_{eff}=\frac{1}{2} \at G_0  f'(G_0)-
f(G_0)-b'G_0\ct\,. \label{exG2} \eeq

In particular by taking $F(G,R)=R$, we  recover the well known
result \cite{oldbrevik}

\be \label{GR14} R_0^2=-\frac{12}{b'}\Lambda_{eff}, \qquad R_0=
-\frac{3}{b'}.\ee

It is important to note that the obtained result (Eq.
(\ref{GR14})) leads us to the inflationary solution (for positive
$H$) obtained first by Starobinsky in Ref. \cite{Sta} using the
renormalized EMT of conformal matter on the right-hand side of
Einstein's equations. This result shows the possibility of
creation of a de Sitter inflationary universe from quantum effects
in the way discussed in \cite{oldbrevik, brevik99, quiroga, EEQJ,
quiroga03, quiroga7}.

Another interesting case is obtained by choosing $f(G)= -\alpha
G^\beta$. As a result one has

\be\label{GR15} \Lambda_{eff}=\frac{1}{2}\alpha G_0^\beta
(1-\beta+b')=\frac{R_0}{4}, \ee

and the solution for $R_0$ has the following form

\be\label{GR16}
R_0=\aq2\alpha\at1-\beta+b'\ct\at\frac{1}{6}\ct^\beta\cq^{\frac{1}{1-2\beta}}.
\ee

Let us now consider several cases for solutions depending on the
values of the parameter $\beta$  in the function $f(G)$.

First of all let us consider the situation where $\beta$ is small.
Then,

\be\label{GR17} R_0\approx2\alpha(1+b').\ee

Neglecting $b'$ in the above result, i.e. if quantum effects are
omitted, we get $R_0\sim\alpha$. This result coincides with that
obtained in \cite{GB-gravity} for modified Gauss - Bonnet gravity
without quantum effects. Furthermore, since $b'$ is negative from
the above results we see that the solution for $R_0$ and therefore
for $\Lambda_{eff}$ is lesser thanks to quantum effects for
$b'>-1$.  For $b'=-1$ the cosmological constant vanishes. In this
way we  conclude that the creation of the inflationary de Sitter
universe occurs only when $b'>-1$.

Let us consider now the case where $\beta=-\frac{1}{2}$. Then one
obtains for $R_0$ the following solution,

\be\label{GR18} R_0=(6)^\frac{1}{4}\sqrt{\alpha\at3+2b'\ct}.\ee

From Eq. (\ref{GR18}) one sees that quantum creation of the de
Sitter universe occurs when $\alpha>0$ only for $b'>-\frac{3}{2}$
and the effective cosmological constant becomes

\be\label{GR19}
\Lambda_{eff}=\frac{1}{4}(6)^\frac{1}{4}\sqrt{\alpha\at3+2b'\ct}.\ee

However one can see that de Sitter universe may also occur even if
$b'< -\frac{3}{2}$ but in such a case $\alpha$ must be negative.

\begin{figure}[h]
\begin{center}
\includegraphics[height=6cm, width= 9cm]{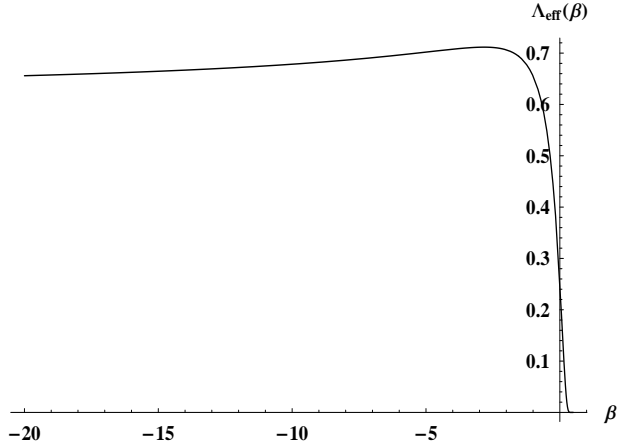}
\caption{Cosmological Constant}
\label{fig1}\end{center}\end{figure}

It is also of interest to see the behavior of the cosmological
function depending on the value of the parameter $\beta$. One
possibility is to evaluate this function taking, for the
parameters $b'$ and $\alpha$,  the values  -0.5 and 1
respectively. As a result, from  figure \ref{fig1}, it is clear
that the cosmological constant is stable for any negative value of
$\beta$ and  vanishes asymptotically when  $\beta$ goes to 0.5.

The combination of quantum effects with modified  gravity may thus
solve the cosmological constant problem bringing the effective
cosmological constant to an extremely small value.

As a final point, let us consider again the case $f(G)=-\alpha
G^\beta$. As noted above, $G_0=\frac{1}{6}R_0^2$ for the de Sitter
solution. If we now assume somewhat more generally that $G$ is
proportional to some power of $R$, we can write the action as
\begin{equation}
S=\int d^4 x \sqrt{-g}\left(R-\alpha' R^{\beta'}+{ \cal L}_m
\right), \label{R}
\end{equation}
$\alpha'$ and $\beta'$ being new constants. Constructing the
equations of motion by the variational procedure, putting
$T^{\mu\nu}$ on the right hand side as usual, one finds that the
covariant divergence of the left-hand side is equal to zero.
Energy-momentum conservation is accordingly a consequence of the
field equations, just as in Einstein's theory. This important
property was demonstrated explicitly by Koivisto
\cite{koivisto06}. The form of action in Eq.~(\ref{R}) was also
considered in \cite{brevik06} in connection with viscous modified
gravity. This action means that we end up with the same
energy-conservation equation as in Einstein's theory:
\begin{equation}
\nabla^{\nu} T_{\mu\nu}=0.
\end{equation}
It means that we are permitted to use the
 generalized form of energy-momentum
tensor corresponding to a viscous fluid, \be T_{\mu\nu}= \rho
U_\mu U_\nu
 +(p-\zeta \theta)h_{\mu\nu},
\ee
 where $\zeta$ is the bulk viscosity, $\theta$ the scalar
expansion, and $h_{\mu\nu}=g_{\mu\nu}+U_{\mu}U_{\nu}$ the
projection tensor (the shear viscosity is omitted). Physically,
this means that we are working to the first order in deviation
from thermal equilibrium. This property of the formalism supports
the  general consistency of the modified gravity theory.

\vspace{3mm}

\noindent {\bf Acknowledgments}

We are grateful to S.D. Odintsov for very helpful discussions. The
research of J.Q.H.  at UTP has been supported by a Professorship
from the Universidad Tecnol\'ogica de Pereira, Colombia.

\end{document}